\documentclass[journal,twoside,web]{ieeecolor}
\usepackage{lcsys}
\usepackage{amsmath,amssymb,amsfonts}
\usepackage{graphicx}
\usepackage{textcomp}
\usepackage{comment}
\usepackage[position=top]{subfig}
\usepackage{multirow}
\usepackage{color,colortbl}
\usepackage[style=ieee]{biblatex}
\usepackage[hidelinks]{hyperref}
\usepackage{url}
\usepackage{doi}
\usepackage{bm}
\usepackage{braket}
\usepackage{mathtools}
\usepackage{dsfont}
\usepackage{physics}

\definecolor{black!15}{gray}{0.9}
\definecolor{black!30}{gray}{0.7}
\definecolor{black!50}{gray}{0.1}

\def\vec#1{\mathbf{#1}}

\def\expectationp#1{\mathds{E}_{\mathbf{P}_\delta}\left[#1\right]}
\def\ss{\mathrm{ss}}
\def\OUT{\mathrm{OUT}}

\newcommand{\out}{\mathrm{out}}

\DeclareMathOperator*{\diffsens}{\ensuremath{\zeta}}

\newtheorem{theorem}{Theorem}
\newenvironment{specialproof}{\textit{Proof:}}{\hfill$\square$}

\title{Analyzing and Unifying Robustness Measures for Excitation Transfer Control in Spin Networks}

\author{S.\,P.\ O'Neil$^{1,*}$ \and I.\ Khalid$^{2,*}$ \and A.\,A.\ Rompokos$^{1}$ \and  C.\,A.\ Weidner$^3$ \and F.\,C.\ Langbein$^2$ \and S.\ Schirmer$^4$ \and E.\,A.\ Jonckheere$^1$%
\thanks{{{$^*$}}first and second author contributed equally to this work}
\thanks{$^1$ Dept of Electrical Engineering, University of Southern California, CA, USA. 
{\tt jonckhee@usc.edu, seanonei@usc.edu, rompokos@usc.edu}
}%
\thanks{$^2$ School of Computer Science and Informatics, Cardiff University, UK. 
{\tt khalidmi@cardiff.ac.uk, frank@langbein.org}
}%
\thanks{$^3$ Quantum Engineering Technologies Laboratories, University of Bristol, UK. 
{\tt c.weidner@bristol.ac.uk}
}%
\thanks{$^4$ Faculty of Science \& Engineering, Physics, Swansea University, UK. 
{\tt s.m.shermer@gmail.com}
}
}

\begin{document}

\maketitle
\thispagestyle{empty}

\begin{abstract}
Recent achievements in quantum control have resulted in advanced techniques for designing controllers for applications in quantum communication, computing, and sensing.  However, the susceptibility of such systems to noise and uncertainties necessitates robust controllers that perform effectively under these conditions to realize the full potential of quantum devices.  The time-domain log-sensitivity and a recently introduced robustness infidelity measure (RIM) are two means to quantify controller robustness in quantum systems.  The former can be found analytically, while the latter requires Monte-Carlo sampling.  In this work, the correlation between the log-sensitivity and the RIM for evaluating the robustness of single excitation transfer fidelity in spin chains and rings in the presence of dephasing is investigated.  We show that the expected differential sensitivity of the error agrees with the differential sensitivity of the RIM, where the expectation is over the error probability distribution.  Statistical analysis also demonstrates that the log-sensitivity and the RIM are linked via the differential sensitivity, and that the differential sensitivity and RIM are highly concordant.  This unification of two means (one analytic and one via sampling) to assess controller robustness in a variety of realistic scenarios provides a first step in unifying various tools to model and assess robustness of quantum controllers.
\end{abstract}


\section{Introduction}\label{sec:intro}
\IEEEPARstart{E}{merging} quantum devices are potentially able to outperform classical computational devices in performing complex and challenging tasks in quantum optics~\cite{scully_zubairy_1997}, quantum cryptography~\cite{Sharbaf_quant_crypto} and quantum machine learning~\cite{Biamonte2017}.  Robust control design is essential to avoid errors in quantum devices that will propagate and amplify as system size scales~\cite{Acharya2023_google}.  

However, the proven techniques of robust control have limited applicability in the control of quantum systems. Standard robust control design and analysis based on small-gain theorem techniques requires closed-loop systems that are well-posed and internally stable~\cite{essentials_of_robust_control}. The marginal stability characteristic of all coherent, ``closed-loop'' quantum systems is thus incompatible with the prerequisites of classical robust control. Further, while the tools of classical control theory are designed to guarantee stability and asymptotic performance, the benefits of quantum technology stem from coherence, a quality that originates with the imaginary-axis poles of quantum systems and rapidly decays with time. As such, quantum control problems such as state transfer or operation of quantum logic gates are generally time-based, depreciating the premium on asymptotic behavior. 

In this letter we consider the task of optimal state transfer in a simple quantum register, focusing on the control paradigm of energy landscape shaping.  The task is formulated as a time-invariant control problem, and optimization techniques are used to identify controllers that yield high fidelity~\cite{Sophie_landscape, Rabitz_landscape}. Uncertainties in the quantum device model and environmental interactions necessitate optimal fidelity controls in the presence of these uncertainties. Various methods to obtain robust and/or optimal controllers exist \cite{koch_survey, mueller2022one, kosut2022}. 
Model-based methods study the problem as an adversarial game between low error and robustness~\cite{PhysRevA.101.052317, https://doi.org/10.48550/arxiv.1911.00789}, whereas model-agnostic methods deploy learning algorithms that rely on input-output measurements to generate robust controllers. 

We examine the correlation between two distinct robustness measures, the time-domain log-sensitivity and a robustness infidelity measure (RIM). Our analysis of the RIM and log-sensitivity is general and not strictly limited to the time-domain or spin systems. Our motivation is to initiate the study of consistent properties between the robustness measures as a first step in development of a unified robustness theory for quantum control. Indeed, while various robustness measures have been used, they have often provided discordant results. Moreover, \emph{reliable} robust control design is important for the successful application of quantum technologies across domains in the noisy real-world setting. In traditional control theory, sensitivity quantifies the performance of a closed-loop system under uncertainties. The time-domain log-sensitivity can be computed analytically~\cite{sean_time_logsens_2022}. Alternatively, the RIM$_p$~\cite{rim_paper} evaluates controller robustness based on the $p$th order Wasserstein distance of the error distribution under uncertainty relative to the ideal distribution. Previous work~\cite{sean_time_logsens_2022} extensively analyzes the log-sensitivity in chains and rings of $N$ particles with structured perturbations and indicates small log-sensitivity is possible for extremely high-fidelity controllers. 

Classically a conflict exists between minimum error and minimum sensitivity of the error quantified as $S(j \omega)+T(j \omega) = I$, where $S$ is the tracking error and $T$ the sensitivity of the error relative to \emph{unstructured} uncertainties~\cite{Safonov_Laub_Hartmann}. Attempts to embed $S$ and $T$ in a single criterion have been proposed, e.g., the ``mixed-sensitivity," and its reformulation for $\mathbf{\Delta}$-\emph{structured} uncertainties as $\mu_{\mathbf{\Delta}} \left( \begin{bmatrix} S^T & T^T \end{bmatrix}^T \right)$~\cite{essentials_of_robust_control}.  
Thus the RIM$_p$ may be viewed as a mixed-sensitivity approach for uncertainties \emph{structured} by their PDFs where both the error (or infidelity) and its robustness (the variance of its PDF) are encoded in a $p$-Wasserstein distance. 

Additionally, the RIM$_1$ is connected to randomized benchmarking that averages the fidelity over the Clifford group used to quantify robustness of quantum circuits by characterizing the error per gate~\cite{randomized_benchmarking}. Moreover, the RIM$_1$ formalizes the already common approach of optimizing for the average infidelity to obtain robust quantum controllers~\cite{khaneja2005optimal}.

The letter is organized as follows. Sec.~\ref{sec:physical_model} introduces the model used throughout this analysis. Sec.~\ref{sec:rob_assess} introduces and unifies the log-sensitivity and the RIM to measure robustness. In Sec.~\ref{sec:results}, we formulate the hypothesis tests to compare the two measures and present our results. Sec.~\ref{sec:conclusion} concludes.

\section{Physical Model} \label{sec:physical_model}
We consider a quantum register comprised of an array of quantum bits. The system can be modeled as a coupled spin system with Hamiltonian
\begin{equation}
   H := \sum_{m\neq n}^N J_{mn} \left(X_n X_m + Y_n Y_m + \kappa Z_n Z_m\right)
   \label{eq:xxcoupling_system_hamiltonian}
\end{equation}
where $N$ is the number of qubits and $X_n$, $Y_n$, $Z_n$ are the $N$-fold Pauli operators acting on the $n$th particle~\cite{Joel_2013}. $J_{mn} = J_{nm}$ denotes the interaction between particles $n$ and $m$ and can be interpreted as the undirected edge between nodes $n$ and $m$ on a graph.  Only 1D spin networks are considered here, with either a chain (linear register) or a ring (quantum router) topology, constraining the couplings to be zero except for $J_{n,n\pm 1}$ (chain) and additionally $J_{1,N}=J_{N,1}$ for rings.  We assume uniform coupling for all non-zero $J$ and $\kappa=0$.  
We further restrict the dynamics to the single excitation subspace and the case where control is achieved by external bias fields that shift the energy levels of particle $n$ by $\Delta_n$, resulting in an effective single excitation subspace Hamiltonian $H_{\ss}$ given by a matrix with diagonal elements $\Delta_n$ and off-diagonal elements $J_{mn}$. The closed system with no interaction with the environment evolves according to $\dot{\rho}(t) = -\tfrac{i}{\hbar} \left[ H_{\ss}, \rho(t)\right]$, where $\rho(t)$ is density operator describing the state of the system~\cite{Blum:1433745,data_set_2_paper}.
 
To study the robustness of a nominally closed quantum system to environmental interaction we introduce a perturbation in the form of dephasing in the Hamiltonian basis. This modifies the evolution of the perturbed state  $\tilde{\rho}(t)$ to
\begin{equation}\label{eq:lindblad_eqn}
  \dot{\tilde{\rho}}(t) = -\frac{i}{\hbar}[H_\ss, \tilde{\rho}(t)] + L(\tilde{\rho}(t))
\end{equation}
where $L(\cdot) = -\tfrac{1}{2}[V, [V, (\cdot)]]$ is the Lindblad decoherence superoperator, and $[\cdot, \cdot]$ is the commutator. We represent the dephasing terms as $V = V^{\dagger} = \sum_{k=1}^N c_{k} \Pi_{k}$, where $\Pi_{k}$ is the projector onto the $k$th shared eigenspace of $H_{\ss}$ and $V$, and $c_k$ is the associated eigenvalue of $V$.  Pre- and post-multiplying~\eqref{eq:lindblad_eqn} by $\Pi_{k}$ and $\Pi_{\ell}$, respectively, and noting that $\{\Pi_k\}$ is a resolution of the identity on $\mathbb{C}^N$, gives
\begin{align}\label{eq:lindblad_sol}
  \tilde{\rho}(t) 
  &= \sum_{k,\ell =1}^{N} e^{-t(i \omega_{k \ell} + \gamma_{k \ell})} \Pi_{k} \rho_{0} \Pi_{\ell},
\end{align}
where $\omega_{k\ell}=(1/\hbar) \left(\lambda_k-\lambda_\ell \right)$ and $\gamma_{k\ell}$ are the decoherence rates and $\rho_0$ is the (known) initial state of the system.

To permit robustness analysis in a linear time-invariant (LTI) framework, we recast~\eqref{eq:lindblad_eqn} as 
\begin{equation}\label{eq: lti_diff}
  \dot{\tilde{r}}(t) = A\tilde{r}(t) + L\tilde{r}(t)
\end{equation}
by expanding~\eqref{eq:lindblad_eqn} with respect to a suitable set $\{\sigma_n\}$ of $N^2$ Hermitian basis matrices for $\mathbb{C}^{N^2}$~\cite{Altafini2012,neat_formula}. Here, $\tilde{r}(t) \in \mathbb{R}^{N^2}$ is the vectorized representation of $\tilde{\rho}(t)$ in the basis $\{\sigma_n\}$ with components $\tilde{r}_k(t) = \Tr(\tilde{\rho}(t)\sigma_k)$.
The matrices $A,L \in \mathbb{R}^{N^2 \times N^2}$ are defined by~\cite{neat_formula}
\begin{subequations}
  \begin{align}
    A_{k \ell} &= \Tr(\frac{i}{\hbar} H_{\ss}  [\sigma_k,\sigma_{\ell}]), \label{eq: mapping-A}\\
    L_{k\ell} &= \frac{1}{\hbar}\Tr(V \sigma_k V \sigma_{\ell}) - \frac{1}{2 \hbar} \Tr(V^2 \left( \sigma_k \sigma_\ell + \sigma_{\ell} \sigma_{k} \right)) \label{eq: mapping-L}.
  \end{align}
\end{subequations}
The solution to~\eqref{eq: lti_diff} is given by $\tilde{r}(t) = e^{t(A + L)}r_0$, where $r_0$ is the expansion of $\rho(0)$. 

\section{Robustness Assessment}\label{sec:rob_assess}

\subsection{Performance and Perturbation Model}

We consider the fidelity error of the excitation transfer from the initial state $\rho(0)$ to a desired output state $\rho_{\out}$ at a read-out time $T$ as the measure of performance. We restrict our analysis to spin rings and chains of size $N=5$ and $N=6$. For chains, we consider transfer from spin $1$ to desired output states $ \text{OUT}=\{ \lfloor N/2 \rfloor +1, N \}$. For rings, we consider transfers from spin $1$ to $\text{OUT} = 2$ through $\lceil N/2 \rceil$. All controllers are optimized to maximize fidelity under varying conditions as described in~\cite{rim_paper,data_set_2_paper}.
We evaluate the nominal fidelity error 
in the LTI formalism as 
$e(T) = 1 - \vec{c}r(T)$ where $\vec{c} \in \mathbb{R}^{1 \times N^2}$ is the transpose of $r_{\OUT}$.

To model the dephasing processes, we use the set of $1000$ dephasing operators specific to spin networks of size $N=5$ or $6$, as employed in~\cite{data_set_2_paper}, normalized and tested to meet the physical complete positivity constraints~\cite{deph_rates}.
We denote this set of dephasing operators by $\{S_{\mu} \} \in \mathbb{R}^{N^2 \times N^2}$ where $\mu$ indexes each dephasing operator and the elements of $S_\mu$ are given by~\eqref{eq: mapping-L} for the LTI representation. 
To modulate the strength of the perturbation we introduce the dimensionless scalar $\delta \in \left[ 0, 0.1 \right]$.

The perturbed trajectory specific to $S_\mu$ and $\delta$ is 
\begin{equation}\label{eq: perturbed_r}
  \tilde{r}(t;S_{\mu},\delta) = e^{t(A + \delta S_{\mu})}r_{0}. 
\end{equation} 
This gives the perturbed performance measure 
\begin{equation}\label{eq: perturbed_err_r}
  \tilde{e}(T;S_{\mu},\delta) = 1 - \vec{c}e^{T(A + \delta S_{\mu})}r_0
\end{equation}
where $\tilde{e}(T;S_{\mu},\delta)$ denotes the error evaluated at time $T$ under the dephasing process $S_{\mu}$ at strength $\delta$.

\subsection{Log-Sensitivity}\label{ssec:log_sens} 

In accordance with~\cite{data_set_2_paper}, we choose the log-sensitivity as one measure of robustness, calculated in two distinct ways: analytically and numerically. In the analytical case we calculate it directly from~\eqref{eq: perturbed_err_r} as in ~\cite{data_set_2_paper}. 
For a given controller and dephasing process we have  
\begin{equation} \label{eq: analytic_log_sens}
  s(S_{\mu},T) = \left. \frac{1}{e(T)}\frac{\partial \tilde{e}(T;S_{\mu},\delta)}{\partial \delta} \right|_{\delta = 0} =  \frac{-1}{e(T)}\vec{c}e^{(TA)}(TS_{\mu})r_0.
\end{equation}
The form of the last term in~\eqref{eq: analytic_log_sens} only holds for this specific dephasing model where $\left[ A,S_{\mu} \right] = 0$. For a given controller, we average the $s(S_\mu,T)$ values to yield $s_{a}(S,T)$ where the subscript $a$ denotes `analytic' and we drop $\mu$ to indicate averaging over the entire set $\{S_{\mu}\}$.

In a complementary manner, we approximate the probability density function $p(\delta,e)$ by sampling the fidelity error $\tilde{e}(T;S_{\mu},\delta)$ of one controller for $1000$ dephasing operators and a range of $\delta$, and calculating a kernel density estimator (KDE).  We quantize the dephasing strength $\delta$ into 1001 steps. For each controller, we then produce a $1001 \times 1000$ array of samples by evaluating $\tilde{e}(T;S_{\mu},\delta)$ at each step of $\delta$ for each dephasing operator $S_{\mu}$.  From this array, we extract the estimated fidelity error distribution through the MATLAB function \texttt{ksdensity}. Selecting a suitable kernel radius for the KDE is crucial to obtain a good estimator.  We leverage the MATLAB function \texttt{smoothingspline} to produce a functional representation of the mean error denoted as $\hat{e}(T;S,\delta)$, where $\mu$ is dropped to indicated that averaging over the dephasing operators has already taken place. We then calculate a numerical derivative of the mean error estimate at $\delta=0$ so that
\begin{equation}
  s_{k}(S,T) = \frac{1}{e(T)}\left. \frac{\partial \hat{e}(T;S,\delta)}{\partial \delta} \right|_{\delta = 0}
\end{equation} 
provides the KDE-based log-sensitivity for a given controller.

\subsection{RIM}\label{ssec:rim} 
Under uncertain dynamics, the fidelity error is a sample drawn from the probability distribution $\mathbf{P}_\delta({\mathbf{e}=e})$ of a random variable $\mathbf{e}$. The subscript signifies that the probability distribution depends on the noise strength $\delta$. The $\mathrm{RIM}_1(\delta)$ (robustness infidelity measure) is the first order Wasserstein distance of $\mathbf{P}_\delta({\mathbf{e}=e})$ from the maximally robust probability distribution, i.e., the Dirac delta distribution at minimum infidelity $0$. 
Note that the $p$-Wasserstein distance between two measures $\mu(dx), \nu(dy)$ is the minimum over all transference plans of the average $p$-moment of $|x-y|$ or cost of transferring $\mu$ to $\nu$~\cite{villani_springer}. We can simplify the $\mathrm{RIM}_1$ as the first raw moment of the error probability distribution~\cite{rim_paper}, 
\begin{equation}\label{eq:rim1_defn}
  \mathrm{RIM}_1(\delta) = \expectationp{e},
\end{equation}
where $\expectationp{\cdot} = \int_{\mathcal{X}} {(\cdot)\mathbf{P}_\delta(\mathbf{e}=e)\,de}$ is the expectation operator w.r.t. the probability distribution of the error $\mathbf{P}_\delta({\mathbf{e}})$ over some appropriate domain $\mathcal{X}$.
The $\mathrm{RIM}_1$ aims to capture both infidelity and robustness in a single measure and extends the infidelity by a noise strength $\delta$. At $\delta=0$, there is no uncertainty so the $\mathrm{RIM}_1$ is just the nominal fidelity error (infidelity) $e(T)$. A further generalization is the $\mathrm{RIM}_p$ as the $p$th order Wasserstein distance can be used, but this is not considered here.

\subsection{Unifying Differential Sensitivity with the RIM}\label{ssec:unify_rim_diffsens} 
We can relate the $\mathrm{RIM}_1$ with the differential sensitivity
\begin{equation}\label{eq:diffsens}
  \diffsens(S_{\mu}, T) = \left.\frac{\partial \tilde{e}(T;S_{\mu},\delta)}{\partial \delta}\right|_{\delta=0}
\end{equation}
by considering the expectation $\expectationp{\diffsens}$.  The dependence of the function $\mathbf{P}_\delta(\mathbf{e}=e)$ and $\diffsens$ on $\delta$ requires careful attention, but for our decoherence noise model, we can use reparametrization~\cite{kingmavae} to write an equivalent expectation operator for our decoherence noise model that isolates the dependence of $\mathbf{P}_\delta(\mathbf{e}=e)$ on $\delta$ to just the error $e$ with a new probability distribution function independent of $\delta$. 

One way to do this is to note that the stochasticity of $\mathbf{e}$ is entirely due to the uncertainty of the dephasing operators $S_{\mu}$, which is represented by the random variable $\mathbf{S}$, with $\delta$ being a deterministic scale parameter.
\begin{theorem}\label{thm:diffsens_rim_relation}
For the decoherence noise model, the expected differential sensitivity is the differential sensitivity of the $\mathrm{RIM}_1$ i.e.
$\mathds{E}_{\mathbf{P}(\mathbf{S})}[\diffsens(S_{\mu}, T)] =  \left.\frac{\partial \mathrm{RIM}_1(\delta)}{\partial \delta}\right|_{\delta=0}$.
\end{theorem}
\begin{specialproof}
We first unpack the differential sensitivity using the definition of the derivative,
\begin{align}
    \nonumber
    \left.\frac{\partial \tilde{e}(T;S_{\mu},\delta)}{\partial \delta}\right|_{\delta=0} 
    &= \left.\lim_{\epsilon\rightarrow0^+}\frac{\tilde{e}(T;S_{\mu},\delta+\epsilon) - \tilde{e}(T;S_{\mu},\delta)}{\epsilon}\right|_{\delta=0} \\
    &=\lim_{\epsilon\rightarrow0^+}\frac{\tilde{e}(T;S_{\mu},\epsilon) - \tilde{e}(T;S_{\mu},0)}{\epsilon}.
    \label{eq:deriv_defn_diffsens}
\end{align}
We apply the expectation operator $\mathds{E}_{\mathbf{P}(\mathbf{S})}\left[ \cdot \right]$ on \eqref{eq:deriv_defn_diffsens} and simplify using the reparametrization trick: $\mathds{E}_{\mathbf{P}(\mathbf{S})}[\cdot] \leftrightarrow \expectationp{\cdot}$,
\begin{align*}
    \mathds{E}_{\mathbf{P}(\mathbf{S})}[\diffsens(S_{\mu},T)] 
    &= \mathds{E}_{\mathbf{P}(\mathbf{S})}\left[\lim_{\epsilon\rightarrow0^+}\frac{\tilde{e}(T;S_{\mu},\epsilon) - \tilde{e}(T;S_{\mu},0)}{\epsilon}\right] \\
    &= \lim_{\epsilon\rightarrow0^+}\frac{\mathds{E}_{\mathbf{P}(\mathbf{S})}\left[\tilde{e}(T;S_{\mu},\epsilon) - \tilde{e}(T;S_{\mu},0)\right]}{\epsilon} \\
    &= \lim_{\epsilon\rightarrow0^+}\frac{\mathds{E}_{\mathbf{P}_\epsilon}\left[\tilde{e}(T;S_{\mu},\epsilon) - \tilde{e}(T;S_{\mu},0)\right]}{\epsilon} \\
    &= \lim_{\epsilon\rightarrow0^+}\frac{\mathrm{RIM}_1(\epsilon) - \mathrm{RIM}_1(0)}{\epsilon} \\
    &= \left.\frac{\partial \mathrm{RIM}_1(\delta)}{\partial \delta}\right|_{\delta=0}.
\end{align*}
Swapping the limit and the expectation in the second line is justified as long as the limit in the mean of the sequence $\{\frac{\tilde{e}(T;S_{\mu},\epsilon) - \tilde{e}(T;S_{\mu},0)}{\epsilon}\}_{\epsilon>0}$ exists.
\end{specialproof}

Note that Thm.~\ref{thm:diffsens_rim_relation} does not necessarily hold in the general case, as removing the dependence on $\delta$ via reparametrization is not always possible.

\section{Results}\label{sec:results}

\subsection{RIM Preprocessing}\label{ssec:RIM_massage}

\begin{figure}
  \centering
  \includegraphics[width=\columnwidth]{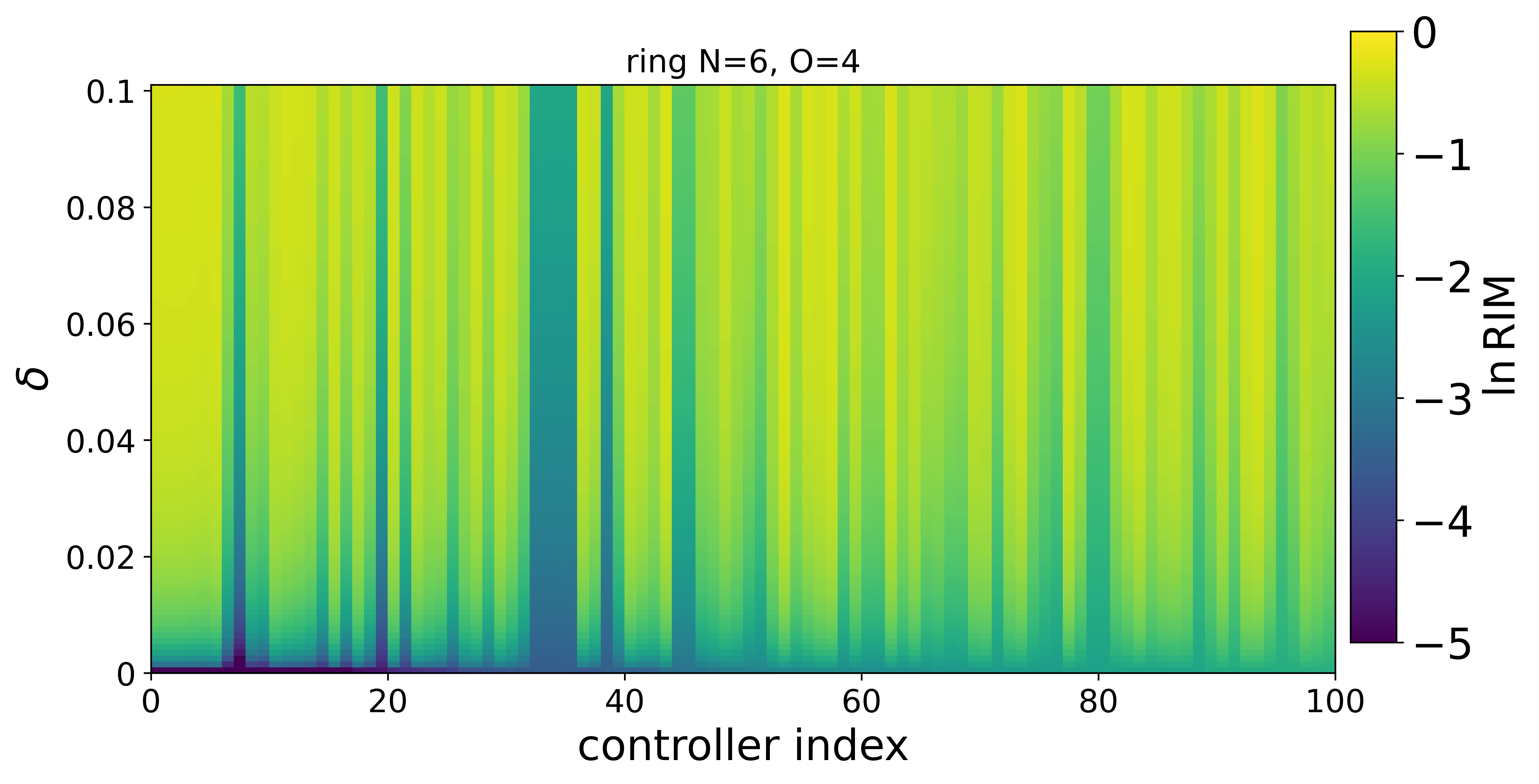}
  \caption{$\mathrm{RIM}_1(\delta)$ for $100$ controllers sorted in increasing order of error (to the right) for the ring spin transfer problem for $N=6$ with $O=4$.}
  \label{fig:rim_cont_heatmap}
\end{figure}
\begin{figure}[t]
  \centering
  \includegraphics[width=0.8\columnwidth]{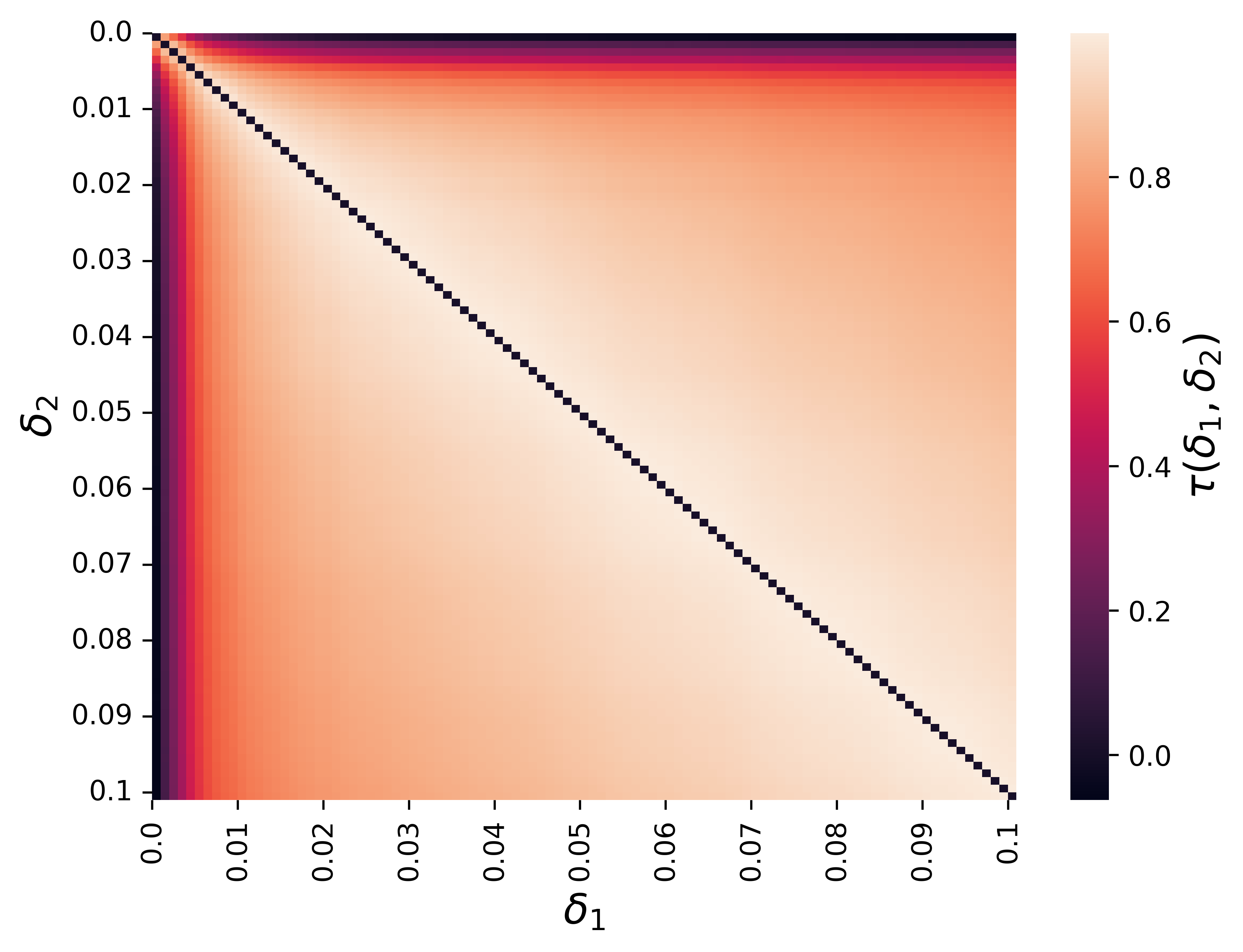}
  \caption{Kendall's tau $\tau(\delta_1, \delta_2)$ heat map showing agreement between $\mathrm{RIM}_1(\delta_1)$ and $\mathrm{RIM}_1(\delta_2)$ measures for $0.1 > \delta_1, \delta_2 > 0.005$ for the ring spin transfer problem for $N=6$ with $O=4$.}
  \label{fig:tauheatmap_to_get_rim_delta}
\end{figure}

To compare the log-sensitivity and $\mathrm{RIM}_1(\delta)$, we need to extract a representative dephasing noise scale $\delta$ to use for $\mathrm{RIM}_1(\delta)$ since the log-sensitivity is independent of $\delta$.
\emph{A priori}, for the two measures $\mathrm{RIM}_1(\delta_1)$ and $\mathrm{RIM}_1(\delta_2)$, for some $\delta_1, \delta_2 \in [0, 0.1]$ noise scale parameters and $\delta_1 \neq \delta_2$, the measure values or controller rank ordering w.r.t. the values do not necessarily coincide or agree. 
We quantify the agreement using rank-correlation analysis via Kendall's tau $\tau(\delta_1, \delta_2)$ for $100$ controllers which are ranked according to their respective $\mathrm{RIM}_1$ values.  For the spin transfer problems considered here, we found that for $\delta_1=0.05$, the rank correlation is strongest (around $>0.8$) for $\delta_2 \in (0.005,0.1)$. Fig.~\ref{fig:rim_cont_heatmap} shows the $\mathrm{RIM}_1(\delta)$, $\delta \in [0, 0.1]$, for $100$ individual controllers sorted in increasing order of error (to the right) for the spin ring transfer problem with $N=6$, $\mathrm{OUT}=4$. Fig.~\ref{fig:tauheatmap_to_get_rim_delta} shows the results of our rank-correlation analysis for the same transfer problem.

\subsection{Hypothesis Test Formulation: Robustness Measure Concordance and Robustness-Performance Trade-off}\label{ssec:hyp_test_formulate} 

Given two forms of the log-sensitivity and the $\mathrm{RIM}_1$, we expect that if all give a trustworfthy measure of robustness, they should show a concordant trend across all controllers for the same problem defined by ring or chain size and transfer.  Based on the fundamental limitation $S(s)+T(s)=I$ of classical feedback control, we also anticipate that the controllers exhibiting good robustness (small log-sensitivity or $\mathrm{RIM}_1$) should have diminished performance (larger fidelity error). To test these hypotheses, we establish a pair of hypothesis tests based the Kendall $\tau$ rank correlation coefficient. To test concordance or discordance of robustness measures we establish one-tailed tests for concordance (right tail for $\tau > 0$) or discordance (left tail for $\tau < 0)$ as
\begin{itemize}
\item $H_0$: no correlation of $s_a(S,T)$, $s_k(S,T)$ and $\mathrm{RIM}_1$;
\item $H_{1+/-}$: positive/negative correlation of same metrics.
\end{itemize}
The rank correlation is computed in a pairwise manner between robustness measures. To test the trend between $e(T)$ and the robustness measures, we establish a second one-tailed test (left tail) for anti-concordance as
\begin{itemize}
\item $H_0$: no correlation of $e(T)$ and $s_a(S,T)$, $s_k(S,T)$, or $\mathrm{RIM}_1$;
\item $H_1$: negative correlation between same metrics.
\end{itemize}
With the combination of ring and chain sizes and transfer problems at our disposal, this provides a total of $36$ test cases for each hypothesis test for chains and $45$ test cases for each hypothesis test for rings. Within each test, we have $100$ samples based on the best (highest fidelity without decoherence) $100$ controllers. 

For each test, we evaluate the statistical significance as 
\begin{equation}
  p_{\tau} = \begin{cases} 
    1 - \Phi(Z_{\tau}),& \text{ for concordance}\\
    \Phi(Z_{\tau}),& \text{ for discordance},
  \end{cases}
\end{equation}
where $Z_{\tau}$ is the Kendall $\tau$ test statistic given as $Z_{\tau} = \tau \left( \sqrt{\tfrac{2(2n+5)}{9n(n-1)}} \right)^{-1}$~\cite{Kendall_tau_significance}, $n=100$ is the number of samples, and $\Phi(\cdot)$ is the normal cumulative distribution function. We set the significance level at a standard $95\%$ so that $\alpha = 0.05$. We reject (accept) the null hypothesis if $p_\tau < \alpha$
($p_\tau \ge \alpha$).

\subsection{Hypothesis Test Results: Robustness Measure Concordance and Robustness-Performance Trade-off}\label{hyp_test_results} 

Applying the hypothesis test to the correlation between the robustness measures provides mixed results. On one hand, across all test cases, $s_a(S,T)$ and $s_k(S,T)$ are highly concordant. However, the correlation between the $\mathrm{RIM}_1$ and either measure of the log-sensitivity provides inconclusive results---neither concordant nor discordant. 
This indicates that utilization of the log-sensitivity and $\mathrm{RIM}_1$ as defined in Sec.~\ref{sec:rob_assess} does not provide an equivalent robustness measure. Table~\ref{table_1} displays the results of the hypothesis test for concordance of the robustness measures for the set of chain controllers, illustrating the inconsistent trends between analytic log-sensitivity and $\mathrm{RIM}_1$ but consistent trend between $s_a(S,T)$ and $s_k(S,T)$.

\begin{table}[t]
  \centering
  \captionsetup{font=small}
  \caption{Excerpt of hypothesis test for concordance of robustness measures for chain controllers optimized with different algorithms $\{A,B,C\}$. Light shading indicates discordant trends. Dark shading indicates failure to reject $H_0$.}\label{table_1}
  \begin{tabular}{|l|c|c|c|c|}\hline
  \cellcolor{black!50} & \multicolumn{2}{c|} { $s_{a}(S,T)$ vs. $s_k(S,T)$}  &  \multicolumn{2}{c|} {$s_{a}(S,T)$ vs. $\mathrm{RIM_1}$} \\ \hline Transfer &  $\tau$ & $p_{\tau}$ & $\tau$    &   $p_{\tau}$ \\ \hline
  N$=5$ OUT$=3$ A & $1.000$ & $0.000$ & $0.201$ &  0.002 \\ \hline
  N$=5$ OUT$=3$ B & $1.000$ &  $0.000$ & $0.487$ &  $0.000$ \\ \hline
  N$=5$ OUT$=3$ C & $1.000$ &  $0.000$ & $0.319$ &  $0.000$ \\ \hline
  N$=5$ OUT$=5$ A & $1.000$ &  $0.000$ & $0.258$ &  $0.000$ \\ \hline
  \cellcolor{black!15}N$=5$ OUT$=5$ B & \cellcolor{black!15}$1.000$ & \cellcolor{black!15}$0.000$ & \cellcolor{black!15}$-0.556$ & \cellcolor{black!15}$0.000$ \\ \hline
  N$=5$ OUT$=5$ C & $1.000$ &  $0.000$ & $0.207$ & $0.001$ \\ \hline
 \cellcolor{black!30}N$=6$ OUT$=4$ A & \cellcolor{black!30}$1.000$ &  \cellcolor{black!30}$0.000$ & \cellcolor{black!30}$0.000$ &  \cellcolor{black!30}$0.498$ \\ \hline
  \cellcolor{black!15}N$=6$ OUT$=4$ B & \cellcolor{black!15}$1.000$ &  \cellcolor{black!15}$0.000$ & \cellcolor{black!15}$-0.214$  & \cellcolor{black!15}$0.001$ \\ \hline
  \cellcolor{black!15}N$=6$ OUT$=4$ C & \cellcolor{black!15}$1.000$ &  \cellcolor{black!15}$0.000$ & \cellcolor{black!15}$-0.202$ &  \cellcolor{black!15}$0.001$ \\ \hline
  \cellcolor{black!15}N$=6$ OUT$=6$ A & \cellcolor{black!15}$1.000$ &  \cellcolor{black!15}$0.000$ & \cellcolor{black!15}$-0.134$ & \cellcolor{black!15}$0.024$ \\ \hline
  \cellcolor{black!15}N$=6$ OUT$=6$ B & \cellcolor{black!15}$1.000$ & \cellcolor{black!15}$0.000$ & \cellcolor{black!15}$-0.639$ &  \cellcolor{black!15}$0.000$ \\ \hline
  \cellcolor{black!30}N$=6$ OUT$=6$ C & \cellcolor{black!30}$1.000$  & \cellcolor{black!30}$0.000$ & \cellcolor{black!30}$0.029$ &  \cellcolor{black!30}$0.336$ \\ \hline
  \end{tabular}
\end{table}

\begin{figure}
  \centering
  \includegraphics[width=1.0\columnwidth]{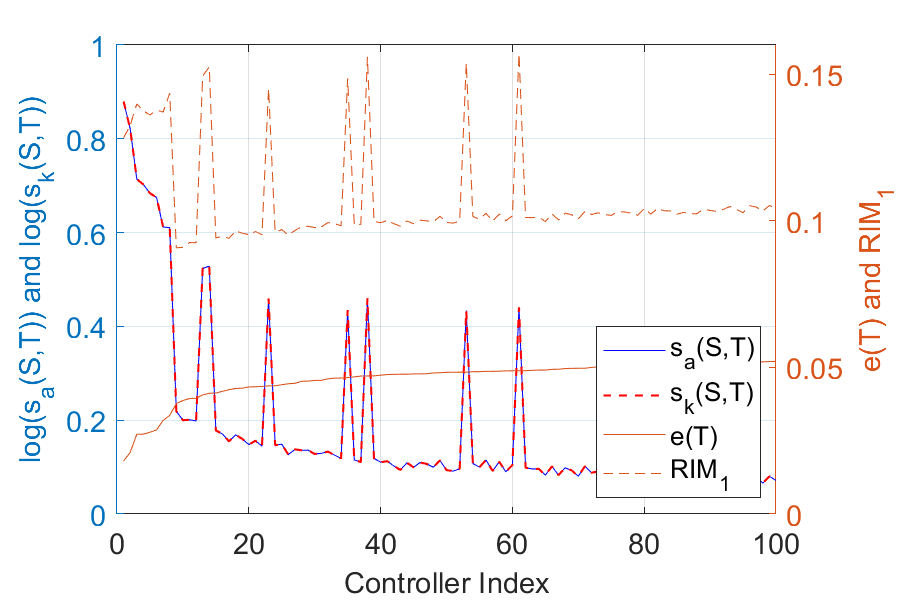}
  \caption{Plot of $s_a(S,T)$, $s_k(S,T)$, and $\mathrm{RIM}_1$ versus controller index (ranked by error) for a $6$-chain, $1 \rightarrow 4$ transfer. The strong correlation between log-sensitivity measures is evident along with the negative correlation between log-sensitivity and $e(T)$. Conversely, we see a concordant trend between $\mathrm{RIM}_1$ and $e(T)$.}\label{fig:chain_6_4_ppo}
\end{figure}

Applying the hypothesis test to the trend between performance and robustness reveals similar, mixed results. For both chain and ring controllers, this trend is highly negative for the log-sensitivity versus $e(T)$, rejecting $H_0$ for $H_1$ in all test cases, indicative of a trade-off between performance and robustness. However, the trend between $\mathrm{RIM}_1$ and $e(T)$ is highly concordant in some cases while anti-concordant in others, a further indicator of dissonance between the robustness measures. Figure~\ref{fig:chain_6_4_ppo} shows a typical plot of the log-sensitivity and $\mathrm{RIM}_1$ versus controller index. Though the trend of log-sensitivity versus $e(T)$ is opposite to that of $\mathrm{RIM}_1$ versus $e(T)$, the plot shows that both measures capture the same ``jumps'', indicating that there is concordance in the ability of each measure to detect the relative robustness between controllers, as predicted by Theorem~\ref{thm:diffsens_rim_relation}.

To explore these relative differences, we consider the relationship between the differential sensitivity $\zeta_{ \{a,k\} }(S,T)$ and the adjusted $\widetilde{\mathrm{RIM}_1} = \mathrm{RIM}_1 - e(T)$ where the nominal error is removed to retain the "spread" of the $\mathrm{RIM_1}$. Reapplying the hypothesis test with the differential sensitivity, calculated both analytically and through the KDE, and $\widetilde{\mathrm{RIM}_1}$ results in strong positive concordance between all three measures for the $45$ ring test cases and $36$ chain test cases with $p$-values near zero. Furthermore, the hypothesis test for anti-concordance between $\zeta_{\{a,k\}}(S,T)$ and the $\widetilde{\mathrm{RIM}_1}$ versus $e(T)$ rejects $H_0$ in favor of $H_1$ in the majority of test cases, while any test cases that do not meet the $\alpha < 0.05$ threshold are the same for all robustness measures. Specifically, for the ring controllers, nine of the $45$ test cases fail to meet the significance threshold, while for the chains, $15$ of the $36$ cases do not meet the threshold. As shown in Table~\ref{table_2},
 however, the data is consistent across the three robustness measures for each controller set, suggesting greater trustworthiness in the differential sensitivity and $\widetilde{\mathrm{RIM}_1}$ to assess robustness.

\begin{table}[t]
  \centering
  \captionsetup{font=small}
  \caption{Excerpt of hypothesis test for trend between differential sensitivity and $\widetilde{\mathrm{RIM}_1}$ versus $e(T)$ for chain controllers optimized with different algorithms $\{A,B,C\}$. Shaded cells indicate failure to reject $H_0$.}\label{table_2}
  \begin{tabular}{|l|c|c|c|c|}\hline
  \cellcolor{black!50} & \multicolumn{2}{c|}{$\zeta_a(S,T)$ vs. $e(T)$} & \multicolumn{2}{c|}{$\widetilde{\mathrm{RIM}_1}$ vs. $e(T)$} \\ \hline Transfer &  $\tau$ & $p_{\tau}$ & $\tau$ & $p_{\tau}$ \\ \hline
  \cellcolor{black!30}N$=5$ OUT$=3$ A & \cellcolor{black!30}$0.0069$ & \cellcolor{black!30}$0.4597$ & \cellcolor{black!30}$-0.0271$ & \cellcolor{black!30}$0.3449$ \\ \hline
  \cellcolor{black!30}N$=5$ OUT$=3$ B & \cellcolor{black!30}$-0.0416$ & \cellcolor{black!30}$0.2698$ & \cellcolor{black!30}$-0.0469$ & \cellcolor{black!30}$0.2448$ \\ \hline
  \cellcolor{black!30}N$=5$ OUT$=3$ C & \cellcolor{black!30}$0.0788$ & \cellcolor{black!30}$0.1227$ & \cellcolor{black!30}$0.0756$ & \cellcolor{black!30}$0.1327$ \\ \hline
  \cellcolor{black!30}N$=5$ OUT$=5$ A & \cellcolor{black!30}$0.0339$ & \cellcolor{black!30}$0.3084$ & \cellcolor{black!30}$-0.0069$ & \cellcolor{black!30}$0.4597$ \\ \hline
  N$=5$ OUT$=5$ B & $-0.1317$ & $0.0261$ & $-0.1426$ & $0.0178$ \\ \hline
  N$=5$ OUT$=5$ C & $-0.1560$ & $0.0107$ & $-0.1754$ & $0.0049$ \\ \hline
  N$=6$ OUT$=4$ A & $-0.3665$ & $0.0000$ & $-0.4097$ & $0.0000$ \\ \hline
  N$=6$ OUT$=4$ B & $-0.2529$ & $0.0001$ & $-0.2590$ & $0.0001$ \\ \hline
  N$=6$ OUT$=4$ C & $-0.2117$ & $0.0009$ & $-0.2246$ & $0.0005$ \\ \hline
  N$=6$ OUT$=6$ A & $-0.2574$ & $0.0001$ & $-0.3220$ & $0.000$ \\ \hline
  N$=6$ OUT$=6$ B & $-0.2178$ & $0.0007$ & $-0.2343$ & $0.0003$ \\ \hline
  \cellcolor{black!30}N$=6$ OUT$=6$ C & \cellcolor{black!30}$0.0238$ & \cellcolor{black!30}$0.3626$ & \cellcolor{black!30}$-0.0008$ & \cellcolor{black!30}$0.4952$\\ \hline
  \end{tabular}
\end{table}

\subsection{On the Differential Sensitivity and Adjusted $\mathrm{RIM}_1$}

Thm.~\ref{thm:diffsens_rim_relation} in Sec.~\ref{ssec:unify_rim_diffsens} shows that the expected differential sensitivity is the differential sensitivity of the $\mathrm{RIM}_1$.  We confirm this with numerical evidence beyond concordance of $\widetilde{\mathrm{RIM}_1}$ and $\zeta_a(S,T)$. Given $\delta$ small enough, a forward difference approximation of $ \left. \tfrac{\partial \mathrm{RIM}_1}{\partial \delta} \right|_{\delta=0}$ shows strong agreement with the value of the $\zeta_{a}(S,T)$. Specifically, quantizing the range of $\delta$ by $1001$ points so that $\delta(n) = 0.0001n$ for $n \in [1,1000]$, the relative error in $\zeta_{a}(S,T)$ and $\left( \mathrm{RIM}_1 - e(T) \right)/\delta(1) = \widetilde{\mathrm{RIM}_1}/\delta(1)$ does not exceed $0.1\%$ across all test cases.

\begin{figure}[t]
\subfloat[Larger perturbation behavior for a $5$-ring $1 \rightarrow 2$ transfer\label{subfig:heat_a}]{\includegraphics[width=0.9\columnwidth]{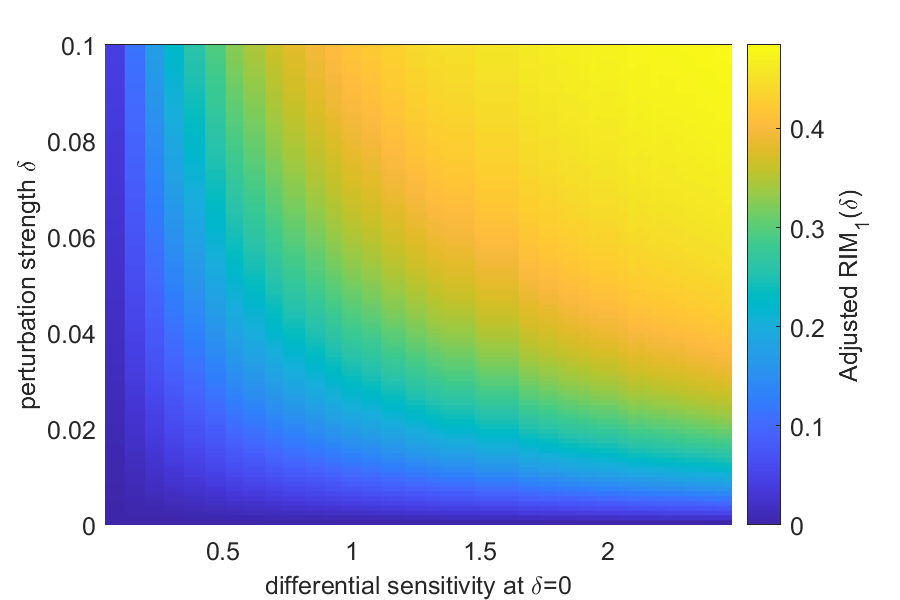}}
\hfill
\subfloat[Larger perturbation behavior for a $6$-ring $1 \rightarrow 4$ transfer \label{subfig:heat_b}]{\includegraphics[width=0.9\columnwidth]{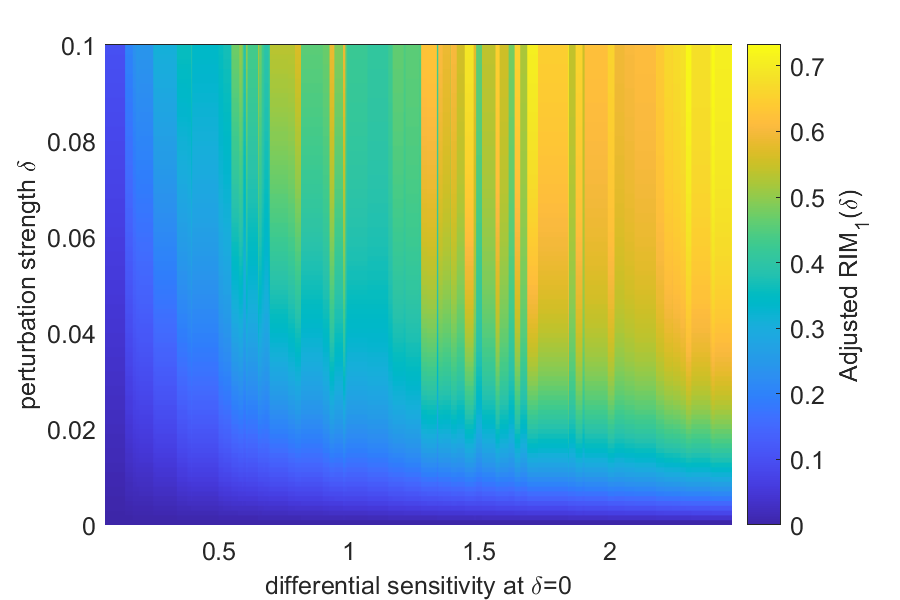}}
\caption{$\widetilde{\mathrm{RIM}_1}$ as a function of $\delta$ compared to $\zeta_{a}(S,T)$.}\label{fig: heat_map}
\end{figure}

Additionally, we see that $\widetilde{\mathrm{RIM}_1}$ has the capability to provide a robustness assessment for values of $\delta$ beyond $\delta=0$ where the differential sensitivity is no longer valid. Fig.~\ref{fig: heat_map} displays characteristic plots of $\widetilde{\mathrm{RIM}_1}$ as a function of $\delta$ ordered by increasing differential sensitivity. Fig.~\ref{subfig:heat_a} shows a characteristic trend of faster increasing $\widetilde{\mathrm{RIM}_1}$ for those controllers with the larger differential sensitivity, suggesting that these controllers display robustness properties at greater perturbation strength in accordance with $\zeta_{a}(S,T)$ at $\delta = 0$. Fig.~\ref{subfig:heat_b} displays the same overall trend but with outliers that indicate the existence of controllers with more global robustness properties that are not captured by the differential sensitivity at $\delta=0$.    

\section{Conclusion}\label{sec:conclusion}

Although the log-sensitivity and $\mathrm{RIM}_1$ have merit as stand-alone measures of robustness, they are not concordant measures. However, they are linked by the differential sensitivity, and we have shown that the expectation of the differential sensitivity over the set of dephasing operators is equivalent to the derivative of $\mathrm{RIM}_1$ at $\delta = 0$. This result is not limited strictly to the time-domain or to spin systems. 
Existing robust control methods can benefit from these measures by using them to post-select synthesized open-loop controllers or directly optimize for controllers that minimize a given robustness measure. Both measures agree locally, near $\delta=0$, but the RIM can be used for a more global (w.r.t. $\delta$) \emph{a posteriori} robustness assessment of controls.

These results are a positive step in unification of robustness measures, but more work is required to make the results more generally applicable. Firstly, the type of perturbations considered must be generalized. Specifically, it is necessary to investigate whether this unification of robustness measures holds under the case of Hamiltonian and/or controller uncertainty simultaneously with dephasing and dissipation. Secondly, investigation of the relationship between higher order differential sensitivity measures and higher orders of the $\mathrm{RIM}$ is necessary to improve higher order robustness of controllers as the number of perturbations under consideration increases.  Finally, a test on physical systems is required to assess how well the proposed robustness measures compare to physically measurable performance in the setting of perturbations and uncertainty. 

\printbibliography
\end{document}